 \documentclass[twocolumn]{aastex62}

\usepackage{graphicx}

\received{2018 September 3}
\revised{2018 October 2}
\accepted{2018 October 3}
\submitjournal{ApJ}

\shorttitle{Off-Limb Flare Loops}
\shortauthors{Jej\v ci\v c et al.}

\begin{document}

\title{High-Density Off-Limb Flare Loops Observed by SDO}

\correspondingauthor{S. Jej\v ci\v c}
\email{sonja.jejcic@guest.arnes.si}

\author{S. Jej\v ci\v c}
\affil{Faculty of Mathematics and Physics, University of Ljubljana, Jadranska 19, 1000 Ljubljana, Slovenia}
\affil{Astronomical Institute, The Czech Academy of Sciences, 25165 Ond\v{r}ejov, Czech Republic}
\author{L. Kleint}
\affiliation{University of Applied Sciences and Arts Northwestern Switzerland, Bahnhofstr. 6, 5210 Windisch, Switzerland}
\affil{Kiepenheuer-Institut f\"{u}r Sonnenphysik, Sch\"{o}neckstr. 6, 79104 Freiburg, Germany}
\author{P. Heinzel}
\affiliation{Astronomical Institute, The Czech Academy of Sciences, 25165 Ond\v{r}ejov, Czech Republic}



\begin{abstract}
The density distribution of flare loops and the mechanisms of their emission in the continuum are still open questions. On September 10, 2017 a prominent loop system appeared during the gradual phase of an X8.2 flare (SOL2017-09-10), visible in all passbands of SDO/AIA and in the white-light continuum of SDO/HMI. We investigate its electron density by taking into account all radiation processes in the flare loops, i.e.\,the Thomson continuum, hydrogen Paschen and Brackett recombination continua, as well as free-free continuum emission. We derive a quadratic 
function of the electron density for a given temperature and effective loop thickness. By absolutely calibrating SDO/HMI intensities, we convert the measured intensities into electron density at each pixel in the loops. For a grid of plausible temperatures between cool (6000 K) and hot (10$^6$ K) 
structures, the electron density is computed for representative effective thicknesses between 200 and 20\,000~km. We obtain a relatively high maximum electron density, about 10$^{13}$~cm$^{-3}$. At such high electron densities, the Thomson continuum is negligible and therefore one would not expect a significant polarization degree in dense loops. We conclude that the Paschen and Brackett recombination continua are dominant in cool flare loops, while the free-free continuum emission is dominant for warmer and hot loops.
\end{abstract}


\section{Introduction}
\label{s_int}

During large eruptive flares, a system of flare loops evolves from the impulsive phase to the often long-lasting
gradual phase \citep{sjm1992}. This is the result of a gradual magnetic reconnection in the corona when the energy is transported
downwards along the reconnected loops and the plasma from strongly heated low atmospheric layers is evaporated.
Due to this process the loops are filled by a hot 10$^6$ - 10$^7$ K plasma, which subsequently cools down. The density distribution of such loops is an open question, but vital to physical models.

Such hot flare loops are now routinely observed e.g. by SDO/AIA \citep{lem12} in selected coronal passbands, while cooler loops cool below transition-region temperatures and finally become visible in chromospheric lines, such as H$\alpha$ \citep[e.g.][]{jing16} or \ion{Mg}{2} \citep[e.g.][]{mik17,lac17}.
These cool
flare loops, often misleadingly called 'post-flare' loops \citep{sve07}, exhibit large downflows, which
is a consequence of the catastrophic cooling. In the
meantime new hot loops form higher in the corona due to gradual reconnection \citep[see][]{sve92}.
This classical scenario corresponds to the so-called CSHKP model \citep{carmichael1964,sturrock1966,hirayama1974,kopppneuman1976} which is widely accepted. 

Although it basically is a 2D model, its generalization to 3D retains similar physics \citep{jan15}. However, the whole process strongly
depends on the efficiency of the reconnection which is, for each event, gradually decreasing with time.
The amount of evaporated plasma directly depends on the amount of energy transported from loop tops down to the transition region and chromosphere. The loop density is thus a crucial parameter needed to understand the temporal evolution of flares, and namely their gradual phases. 

At the beginning the cooling process may be dominated by conduction, while later on the radiative cooling takes over, which is proportional to density squared (or emission measure). In hot loops, the electron density or emission
measure can be diagnosed using various coronal lines, while cool loops with downflowing blobs pose a
more difficult problem. Being detected in cool chromospheric lines, their spectral diagnostics require
complex non-LTE radiative transfer performed for moving structures illuminated by the surrounding
atmosphere. Downward motions cause the so-called Doppler brightening (e.g. in the H$\alpha$ line) or
Doppler dimming  \citep[for \ion{Mg}{2} see][]{mik17} which must be properly taken into account in order
to accurately derive the electron densities. For static loops (e.g. loop tops), \citet{hk87}
derived electron densities of the order of 10$^{12}$ cm$^{-3}$ for H$\alpha$ loops visible in
absorption against the solar disk, while at higher densities the loops may appear in emission. 

Recent observation using the SDO/HMI instrument  revealed flare loops above the limb, surprisingly well detectable in the visible continuum. First detections were reported by \citet{oliv14} and \citet{shil14} after X-class flares. These white-light (WL) loops reached heights of more than 10$^4$ km and the authors suggested that their brightness is due to the Thomson scattering of the incident photospheric radiation on loop electrons, with a possible thermal component (free-bound and free-free). They also used HMI's linear polarization to derive the electron density from the Thomson-scattering component.

On the disk, HMI was used to detect detect ribbons of many WL flares  \citep[e.g.][]{kuh16}, assuming that the outermost HMI channel detects the visible continuum and not the \ion{Fe}{1} line emission during the flare, which was shown to be the case, even though the absolute value of the enhancement may be misrepresented with this method \citep{sva18}.

However, in the off-limb structures the photospheric \ion{Fe}{1} line is not seen and we seem to detect only the visible continuum around that wavelength. We would like to clearly distinguish the observations of chromospheric footpoints of otherwise hot flare loops \citep[e.g.,][]{kru15, hei17} and the full WL-loops of \citet{oliv14} and \citet{shil14}, and of this paper. Here we will analyze a very bright loop system that was detected during the gradual phase of the X8.2 limb flare on September 10, 2017. After calibrating the HMI images, we derive plausible ranges of the electron densities for this event, considering quantitatively all relevant emission mechanisms.

The paper is organized as follows. In Section~\ref{s-obs} we present the SDO/HMI observations and data processing, Section~\ref{s-multi} discusses the multi-thermal nature of flare loops. Section~\ref{s-rad} details all considered emission mechanisms and 
develops the new diagnostics technique for the electron density determination, while 
Section~\ref{s-ne} presents the results of our diagnostics. Finally, Section~\ref{s-con} contains a discussion and conclusions.

\section{Observations and data processings of the loops}
        \label{s-obs}

 \begin{figure*}[tb] 
  \centering 
   \includegraphics[width=.8\textwidth]{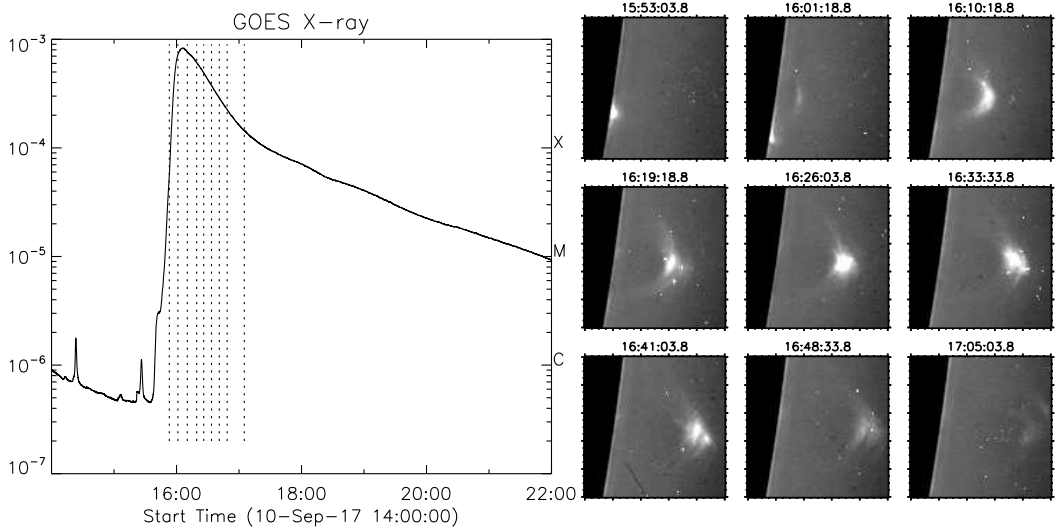}
   \caption{Left: The GOES X-ray 1--8 \AA\ flux (solid line) showing the X8.2 flare. The vertical dotted lines indicate the times of the panels on the right, which show the HMI continuum images of the evolving loop system. The off-limb intensity was enhanced for the display by dividing the regular HMI images by an exponential function and by setting the disk values to zero. A movie showing the full evolution is available online [20170910\_wl\_exp.mp4].}
           \label{f-goes}
  \end{figure*}
  
\begin{figure*}    
  \centering 
   \includegraphics[width=.95\textwidth]{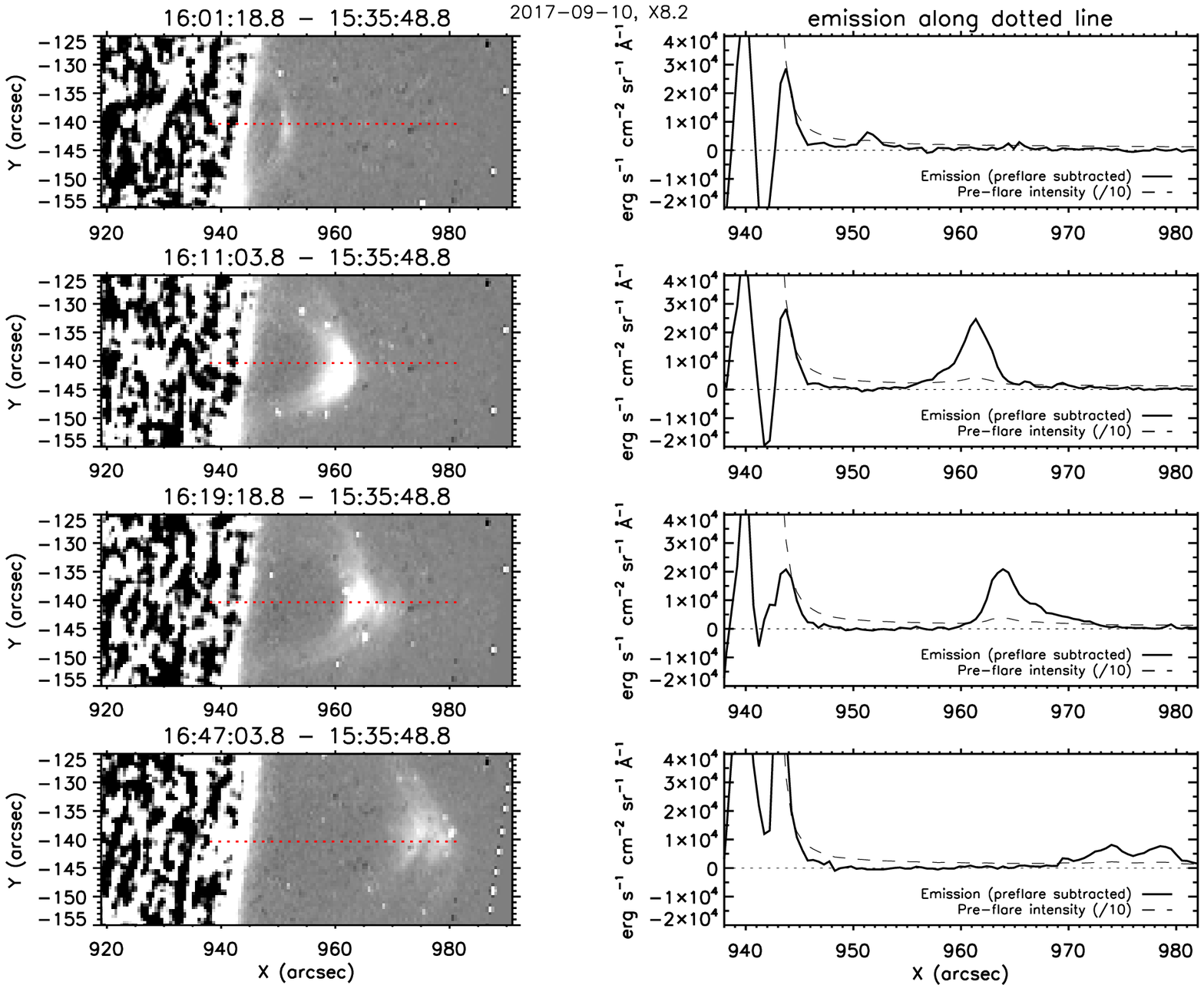}
\caption{Temporal evolution of SDO/HMI WL loops and their intensity. The left panel shows the HMI WL image with a cut through the flare loop (dotted red line). The 
solid line in the right panel shows calibrated intensities with the pre-flare subtracted in CGS units along the marked red dotted line. The dashed line shows the pre-flare intensity divided by 10.}
\label{f-loopint}
\end{figure*}

Active region (AR) 12673 erupted near the west solar limb on September 10, 2017 with its maximum X-ray emission at 16:06 UT as a
strong X8.2 limb flare with well visible arcades of flare loops during its gradual phase. Its GOES X-ray plot and several snapshots of WL data from SDO/HMI are shown in Figure~\ref{f-goes}. We use the continuum channel of SDO/HMI (hmi.Ic\_45s), which is outside of the \ion{Fe}{1} line at 6173 \AA\ \citep{Scherrer2012}. The loop system was visible for more than one hour in the WL images of SDO/HMI.

We performed an absolute calibration of HMI intensities by taking their disk center value, which is about $5.63 \times 10^4$ counts and assigning it the  continuum value from the atlas of \citet{neckel} at 6173 \AA, which is $0.315 \times 10^7$ erg s$^{-1}$ cm$^{-2}$ sr$^{-1}$ \AA$^{-1}$. We applied this conversion factor to the off-limb intensities. We also removed a large fraction of cosmic rays by checking if their intensity at a given time step exceeds three standard deviations of 134 time steps. If a pixel exceeded this threshold at 1 or 2 consecutive time steps, it was flagged as cosmic ray pixel and its value was substituted with the median value from the 2 previous and 2 posterior time steps. This rather conservative approach made sure that we did not filter long-lasting events (more than 2 time steps), but some cosmic rays remain as can be seen in the Figures.

The left panel of Figure~\ref{f-loopint} shows the temporal evolution of the flare loops during the gradual phase between 16:01 and 16:47 UT when the WL continuum emission enhancement is observed by HMI. In the right column we can see the variation 
of the specific intensity of the WL continuum radiation along the 
horizontal cut marked by a red dotted line through the flare loop after subtraction of pre-flare images. We later convert these intensities into  electron densities and can thus also determine the maximum electron density $n_{\rm e}$ in the flare loops.

We further coaligned the SDO/HMI data with all EUV channels of the 
SDO/AIA data. These channels cover the temperature range between 5 
$\times$ 10$^{4}$ and 2 $\times$ 10$^{7}$~K by observing the transition region and corona \citep{lem12}. The spatial resolution of AIA is 1\farcs2, while for SDO/HMI it is 1\arcsec\ and the temporal resolution is 12 s for AIA images and 45 s for HMI observations.

\section{Multi-thermal loops seen by SDO/AIA} 
        \label{s-multi}

\begin{figure*}    
 \centering 
   \includegraphics[width=.95\textwidth]{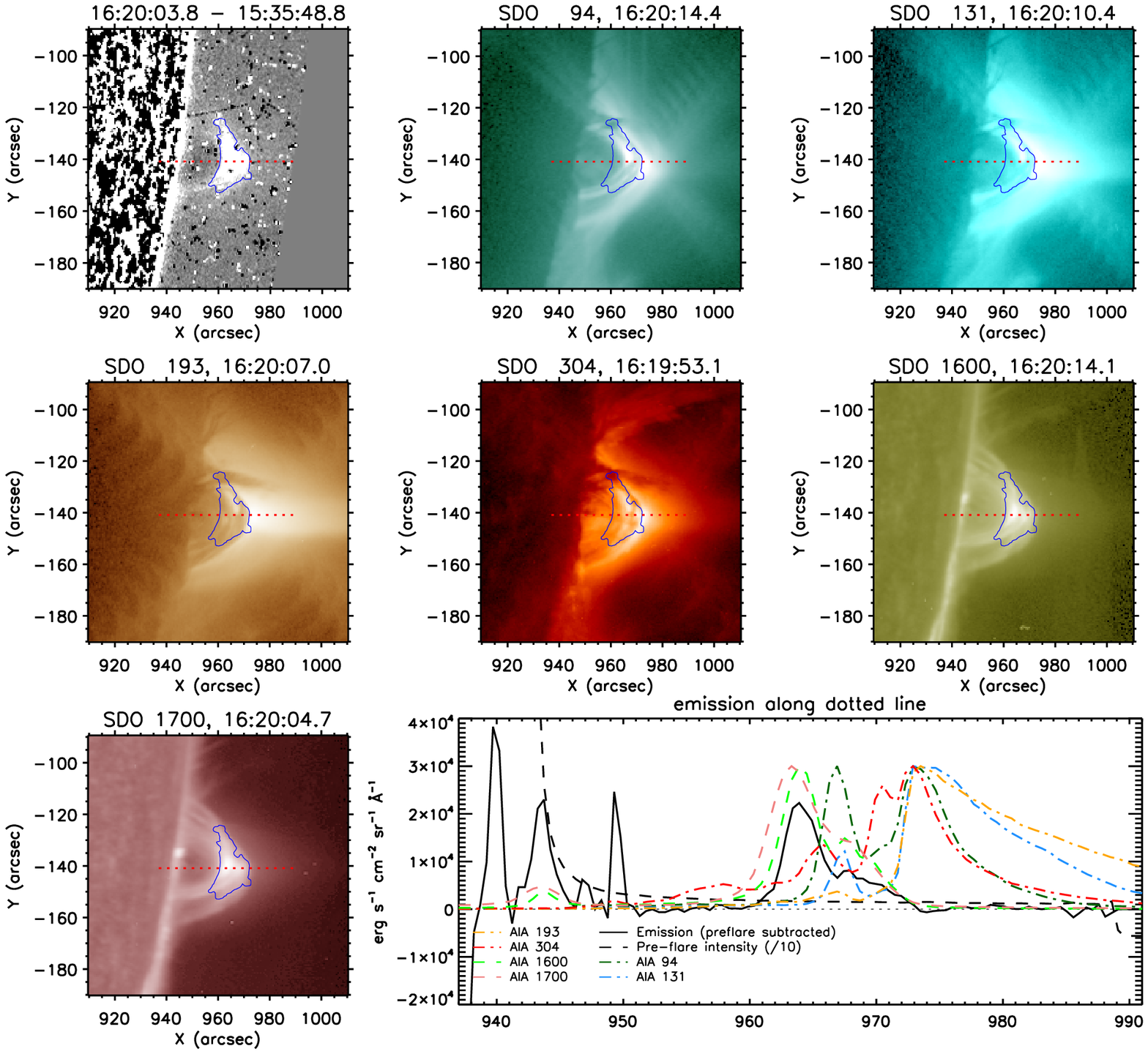}
\caption{An example of flare loops of AR 12673 on the west limb on September 10, 2017 at 16:20:03.8 UT in SDO/HMI (upper left plot, in this case without a correction for cosmic rays) and various SDO/AIA passbands (labeled in their titles). In the HMI panel a pre-flare image from 15:35:48.8 UT was subtracted.
HMI loop contours (marked in blue) are overlaid on all images.
The emission along the red dotted line is shown for all channels in the bottom right panel. The absolute units are valid for SDO/HMI, all AIA wavelengths were scaled to have their maximum at the same level. It is visible that the loop height in HMI coincides with AIA 1600, while the AIA 1700 loop is slightly lower and the loop heights of all other AIA passbands peak higher.}
\label{f-aia}
\end{figure*}

Apart from the SDO/HMI WL evolution, we also examine the behavior of the flare loops in various SDO/AIA channels. The AIA images clearly show a multi-thermal nature of the observed loop arcade and we can deduce spatial correlations between loops in different AIA channels and those detected by HMI in the WL. In this section we provide a qualitative description of the multi-thermal behavior and discuss possible mechanisms responsible for the formation of AIA diagnostics. 

Most AIA channels and a pre-flare subtracted HMI image are shown in Figure~\ref{f-aia}. In this HMI image, the cosmic rays were not removed on purpose to show their prevalence. We also plot the intensity variations along one cut through the loop system (red dotted line in Figure~\ref{f-aia}). The absolute intensity on the y-axis refers only to HMI, all AIA signals were scaled to have their maximum at an arbitrary value.

\subsection{AIA 193, 131 and 94}

The AIA 193 channel shows both hot loop emission in the \ion{Fe}{12} and \ion{Fe}{24} lines, as well as cool loops at chromospheric temperatures seen as dark absorbing features, also reported by \citet{son16}. In this case, the cool loops in front of the hot ones absorb the EUV radiation emitted by iron lines, and the absorption process corresponds to photoionization of the hydrogen and helium by background EUV photons. The dominant process is the photoionization of helium and since the cross-sections of \ion{He}{1} and \ion{He}{2} at 193 \AA\ are about the same, we do not need to consider
\ion{He}{1} and \ion{He}{2} separately \citep{anz05}. The \ion{He}{1} and \ion{He}{2} photoionization continua start at 504 \AA\ and 228 \AA, respectively. 

We clearly see that the intensity depression in the 193
channel (yellow dash-dotted line in Fig.~\ref{f-aia}), downwards from roughly X=971\arcsec, correlates well with the position of the WL loops from HMI. The fact that hot 193 loops are located above those of WL is consistent with 
the standard scenario of gradual reconnection where the reconnected hot loops appear higher and higher, but at a given height they cool down and appear gradually at lower and lower temperatures \citep[for a brief overview of flare loops characteristics see e.g.][]{mik17}. However, the loop arcade studied in this paper seems to be aligned along the line of sight and due to the projection of differently inclined loops we may partially see also the hot loops behind the cool ones, with their emission attenuated by photoionization in cool loops. If the cool loops are optically thick in the \ion{He}{1}-\ion{He}{2} continua (plus some opacity from the \ion{H}{1} Lyman continuum), the hot background loops will not be visible in 193, but will definitely contribute to the HMI WL continuum emission which is optically thin and thus covers the whole arcade along the line of sight (see below). A presence of hotter loops behind the cool ones seems to be indicated by a perfect co-alignment between the HMI WL loops, the dark 193 loops and loops seen in emission in the AIA 1600 and 1700 channels. The other two AIA channels, 131 and 94, exhibit a similar behavior as 193 one, but the absorption is decreasing with the decreasing wavelength and the emission is due to different iron ions \citep{lem12}.

\subsection{AIA 1600 and 1700}

We believe that the AIA 1600 channel loops are mainly due to \ion{C}{4} line emission at temperatures of the order of 10$^5$ K meaning that cool (10$^4$ K) loops and hotter \ion{C}{4} loops are located at similar heights because the time to cool \ion{C}{4} loops down to cool loops is very short, especially at the rather high electron densities we derive in this study - see the cartoon of gradual reconnection and cooling in \citet{sch96}. 
AIA 1700 channel contains mainly cooler lines like \ion{C}{1} and \ion{He}{2} \citep{sim18} which might explain that the emission peak is slightly shifted towards the lower heights.

\subsection{AIA 304}

AIA 304 images show a more complex morphology where a combination of bright and dark loops is visible. In our opinion, this is a mixture
of cooler loops at the temperature of \ion{He}{2} 304 \AA\ line formation (around 5 $\times 10^4$ K) which is seen in emission above the limb, but because of the large opacity in the \ion{He}{2} line some forefront loops can obscure this emission producing dark absorbing features. However, a striking feature is the extension of 304 loops well above those at 1600 \AA. This is somewhat difficult to explain because \ion{He}{2} 304 \AA\ loops cannot form at temperatures higher than the 1600 loops with \ion{C}{4} and thus, according to the above-mentioned reconnection scenario, they should not be located higher than 1600 loops. However, the 304 AIA channel may be contaminated by the 
nearby \ion{Si}{11} line at 303.3 \AA\ and it can produce emission at altitudes higher than those of the 1600 channel, but still somewhat lower than the 193 \ion{Fe}{12} loops. 

It is not the aim of this study to perform a quantitative analysis of emissions or absorptions in all these AIA channels and we provide this discussion just to relate the HMI WL loops to structures seen by SDO/AIA. However, in the future it would be important
to analyze these height variations in detail, for example to estimate densities from the EUV absorptions, or to deduce the (differential) emission measure in hot loops. Such parameters could be compared with our findings for HMI loops.

\section{Continuum radiation processes in flare loops}
         \label{s-rad}

\begin{table} \small 
\centering
\caption{Parameters of WL flare loops: time $t$, height above the solar disk, peak of the specific intensity of the WL continuum radiation, dilution factor $W(H, \nu)$, and diluted mean intensity of the incident radiation from the solar disk $J(\nu)$. 
Here cgs represents the units erg~s$^{-1}$~cm$^{-2}$~sr$^{-1}$. 
Note that the last image has two peaks and the values are shown for both of them: the last two rows for the left and right peak, respectively.}
\label{t-par}
\begin{tabular}{ccccc}
\hline
\hline
{\it t} & {\it H} & {\it I$_{\rm WL}$} & {\it W(H, $\nu$)}&{\it J($\nu$)} \\
 (UT) & (km) & (cgs~\AA$^{-1}$)& &(10$^{-5}$~~cgs~Hz$^{-1}$)  \\
\hline
 16:01:18:8 & 5500  & 6600 & 0.329 & 1.266 \\
 16:11:03.8 & 12\,500 & 24\,000 & 0.313 & 1.204 \\
 16:19:18.8 & 14\,500 & 20\,000 &0.308  & 1.188 \\
 16:47:03.8 & 21\,500 & 7800 & 0.294 & 1.132 \\
 16:47:03.8 & 25\,000 & 7700 & 0.287 & 1.105 \\
\hline
\end{tabular}
\end{table}

WL continuum emission in flare loops observed off the limb is mainly due to four different mechanisms: 
i) Thomson scattering of the incident solar radiation on flare loop electrons, 
ii) hydrogen Paschen recombination continuum (i.e., protons capture free  thermal electrons) with the
continuum head at 8204~\AA~, 
iii) hydrogen Brackett recombination continuum with the continuum head at 14584~\AA, and finally 
iv) hydrogen free-free continuum emission due to energy losses of free thermal electrons in the electric field of protons. Here we
neglect higher hydrogen recombination continua and other continuum sources.
Below we present the explicit forms of emissivities for all these processes \citep[see also][]{hub15,hei17,hei18}. 

\subsection{Optically-thin loops}

The specific intensity of optically-thin continuum radiation is generally written as 
\begin{equation}
I_{\rm WL}(\nu)= \eta(\nu)~D_{\rm eff} \ , 
\label{e-ivl}
\end{equation} 
where $D_{\rm eff}$ is the effective thickness, $\eta(\nu)$ is the emissivity and $\nu$ the frequency, in our case corresponding to the wavelength around the HMI \ion{Fe}{1} line at 6173~\AA. 

The Thomson continuum emissivity is expressed as 
\begin{equation}
\eta^{\rm{Th}}(\nu) = n_{\rm e}~\sigma_{\rm T}~J(\nu) \ ,
\label{e-th}
\end{equation}
where $\sigma_{\rm T} = 6.65 \times 10^{-25}$~cm$^2$ is the absorption cross-section for Thomson scattering 
and $J(\nu)$ is the intensity  of radiation emitted from the 
solar disk center multiplied by a dilution factor $W(H, \nu)$ which takes into account center-to-limb continuum variation and depends on 
the loop height $H$ and frequency.  $W(H, \nu)$ and
$J(\nu)$  are shown in Table~\ref{t-par} and are computed according to \citet{jej09}. 

The Paschen and Brackett continuum emissivity is written as
\begin{equation}
\eta^{i}(\nu) = n_{\rm e}~n_{\rm p}~F_i(\nu,T) \ ,
\label{e-pabr}
\end{equation} 
where the principal quantum number is $i = 3$ for the Paschen (Pa) continuum and $i = 4$ for the Brackett (Br) continuum. 
$n_{\rm e}$ and $n_{\rm p}$ are the electron and proton densities, respectively, and $T$ is the kinetic
temperature of the loop. The function $F_i(\nu,T)$ takes the form \citep{hei17}
\begin{eqnarray}
F_i(\nu,T) &=& 1.1658 \times 10^{14}~g_{\rm bf}(i, \nu)~T^{-3/2}~B_{\rm \nu}(T) \nonumber\\
&\times& e^{h\nu_i/kT}~(1 - e^{- h\nu/kT})~(i \nu)^{-3} \,  .
\label{e-f}
\end{eqnarray}
Here $h$ and $k$ are Planck and Boltzmann constants, respectively, $g_{\rm bf}(i, \nu)$ is the bound-free Gaunt factor and 
$B_{\nu}(T)$ is  the Planck function. $\nu_i$ is the frequency at the respective continuum head. 
The Paschen and Brackett bound-free Gaunt factors at 6173~\AA~are $g_{\rm bf}(3, \nu)$= 0.942 and $g_{\rm bf}(4, \nu)$= 0.998, respectively \citep{mih67}.

The hydrogen free-free continuum emissivity is simply related to the Paschen emissivity as \citep{hei17}
\begin{equation}
\eta^{\rm ff}(\nu) = 8.546 \times 10^{-5}~\frac{g_{\rm ff}(\nu,T)}{g_{\rm bf}(3, \nu)} T e^{-h\nu_3/kT} \times \eta^{\rm Pa}(\nu) \,  .
\label{e-ff}
\end{equation} 
Here $g_{\rm ff}(\nu,T)$ is the free-free Gaunt factor \citep[see][Table~1]{ber56}.

The total WL radiation intensity of an optically thin loop takes into account all four processes, i.e.
\begin{equation}
I_{\rm WL}(\nu) = I^{\rm Th}(\nu) + I^{\rm Pa}(\nu) + I^{\rm Br}(\nu) + I^{\rm ff}(\nu) \ .
\label{e-4pr}
\end{equation}
Equation~(\ref{e-4pr}) can be written using Equations~(\ref{e-ivl}) - (\ref{e-ff}) and assuming  
pure hydrogen plasma with $n_{\rm e} = n_{\rm p}$ as
\begin{eqnarray}
I_{\rm WL}(\nu) &=& n_{\rm e} \sigma_{\rm T} J(\nu) D_{\rm eff} + n_{\rm e}^{2} F_3(\nu, T) D_{\rm eff} \nonumber\\
&\times& (1 + 8.546 \times 10^{-5} \frac{g_{\rm ff}(\nu,T)}{g_{\rm bf}(3, \nu)} T e^{-h\nu_3/kT}) \nonumber\\
&+&  n_{\rm e}^{2} F_4(\nu, T) D_{\rm eff} \,  .
\label{e-all}
\end{eqnarray}
We thus obtained a {\em quadratic equation} to be solved for $n_{\rm e}$, at a given frequency, temperature, and effective 
thickness and for a given (measured) intensity of the WL radiation. 
Note that the second line in Equation~(\ref{e-all}) shows the relative importance of the Paschen and free-free continua.

\begin{figure*} [tbh]   
 \centering 
   \includegraphics[width=\textwidth]{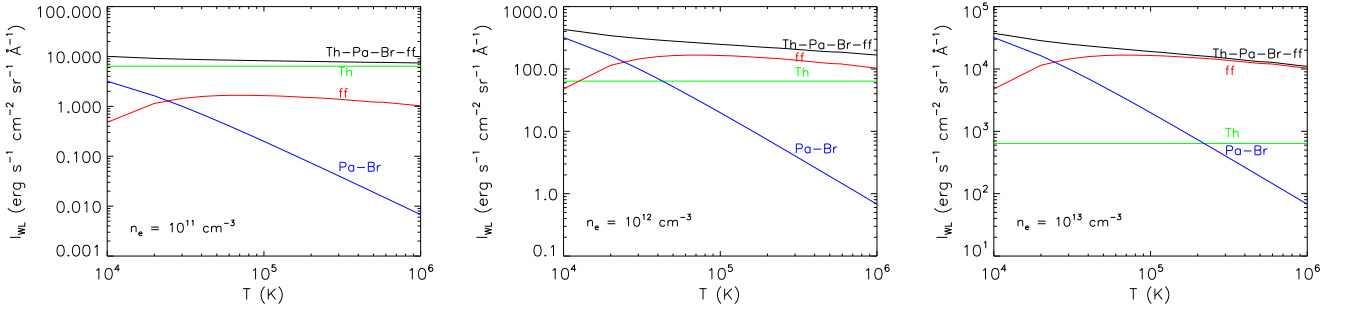}
\caption{Contribution of individual processes (Thomson, free-bound, free-free) to the flare loop WL emission as a 
function of temperature at $H$= 10$^{4}$~km, $D_{\rm eff}$ = 1000~km and for $n_{\rm e}$ = 10$^{11}$~cm$^{-3}$ 
({\it left panel}), $n_{\rm e}$ = 10$^{12}$~cm$^{-3}$ 
({\it middle panel}), and $n_{\rm e}$ = 10$^{13}$~cm$^{-3}$ ({\it right panel}). The Thomson continuum only dominates for low electron densities. At high densities, the Paschen and Brackett continua dominate at lower temperatures, and the free-free emission at high temperatures. }
\label{f-modne}
\end{figure*}

\begin{figure} 
\centering
\includegraphics[width=0.425\textwidth]{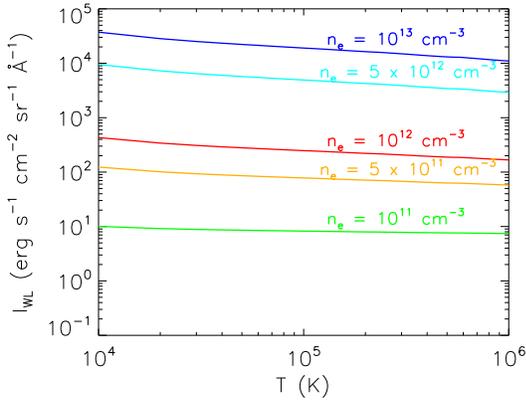}
\caption{Computed WL radiation intensity as a function of temperature for a selected range of electron densities at $D_{\rm eff}$ = 1000~km.
 Note that all four processes (Thomson, Paschen, Brackett and free-free) are taken into account for the WL emission.}
\label{f-ica}
\end{figure} 

\subsection{Relative contribution of the different emission mechanisms}

The computed WL emission for optically thin structures as a function of temperature is shown in Figure~\ref{f-modne} for 
$n_{\rm e}$ equal to  10$^{11}$, 10$^{12}$, and 10$^{13}$~cm$^{-3}$ and a characteristic $D_{\rm eff}$ = 1000~km. At high electron densities, the Thomson continuum is completely negligible compared to the total WL emission. The Paschen and Brackett continua are dominant only at lower temperatures up to about 2.5 $\times$ 10$^4$~K. 
At higher temperatures, the free-free continuum becomes dominant. Therefore, the flare loop WL emission 
can be due to both cool as well as  hot loop structures.

To show the dependence of the electron density diagnostics on temperature, we computed the WL emission from Equation~(\ref{e-all}) 
for a range of temperatures between 10$^4$ and 10$^6$~K and for five different electron densities 10$^{11}$, 5 $\times$ 10$^{11}$, 10$^{12}$, 5 $\times$ 10$^{12}$ and 10$^{13}$~cm$^{-3}$ at $D_{\rm eff}$ = 1000~km. Figure~\ref{f-ica} shows the flare loop WL radiation intensity at given temperature and electron density, computed by adding together all processes. It is visible that the temperature only has a minor influence on the electron density  determination for a given intensity.

\section{Electron density in flare loops}
        \label{s-ne}

\begin{figure*}    
\centering
\includegraphics[width=.81\textwidth]{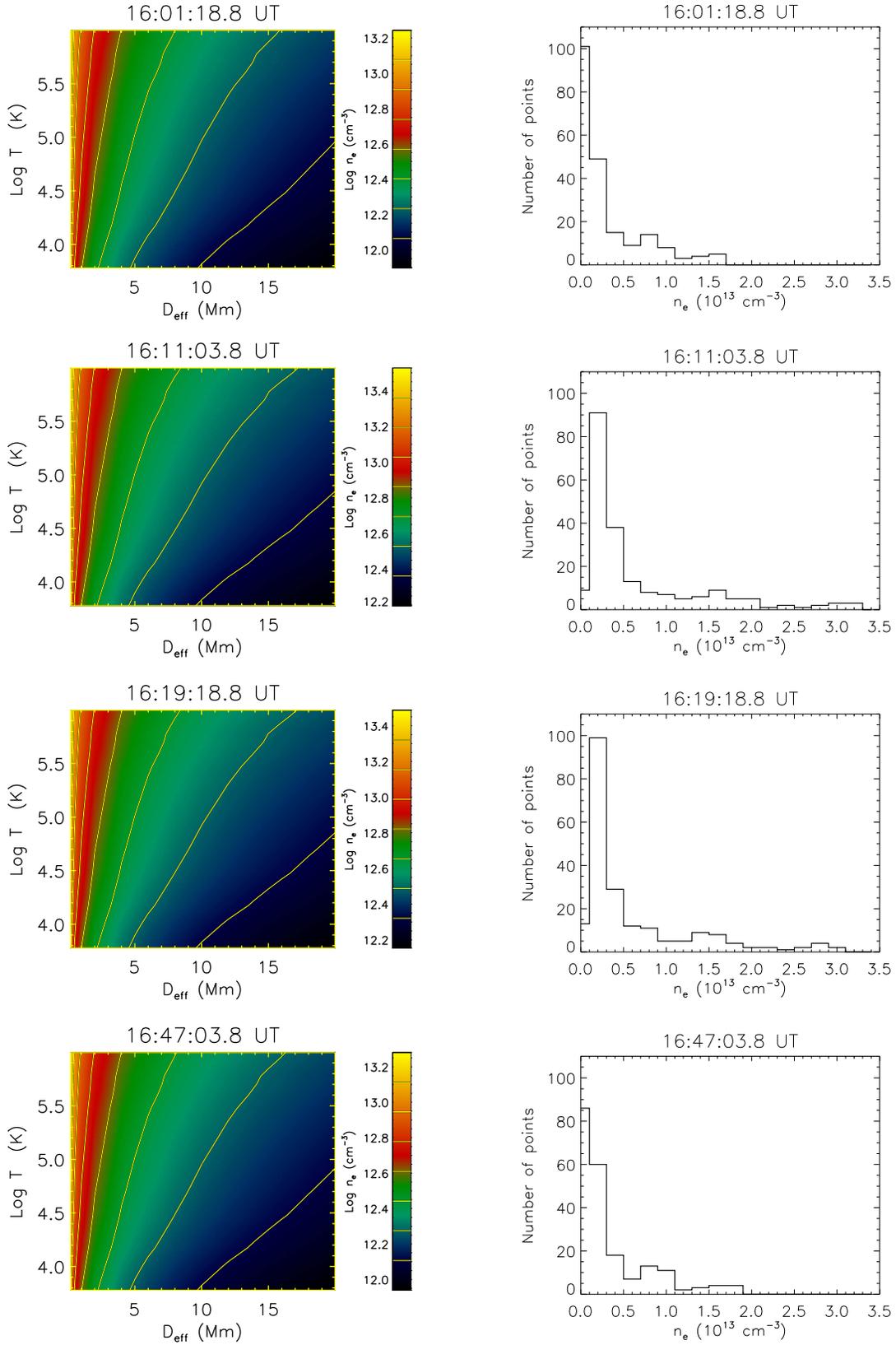}
\vspace{0mm}
\caption{Temporal evolution of contour plots of electron density as a function of effective thickness and temperature for the maximum intensity at a given time ({\it left panel}) together with the distribution of electron density ({\it right panel}) for our grid of 208 models that were constructed from sampling the parameter space shown in the left panel. }
\label{f-net}
\end{figure*}

\begin{figure}    
\centering
\includegraphics[width=.48\textwidth]{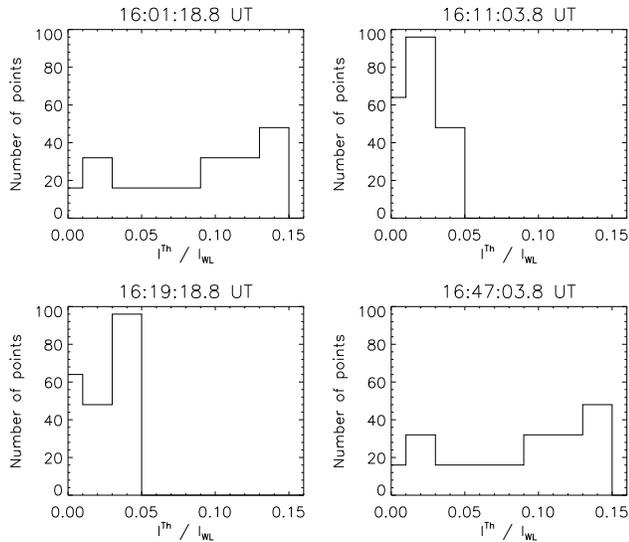}
\caption{Distribution of the ratio of Thomson continuum radiation to the total WL radiation intensity for selected time steps of the selected pixel in the observations.}
\label{f-his}
\end{figure}

\begin{figure*}    
\centering
\hspace{-0.5cm}
\includegraphics[width=1.02\textwidth]{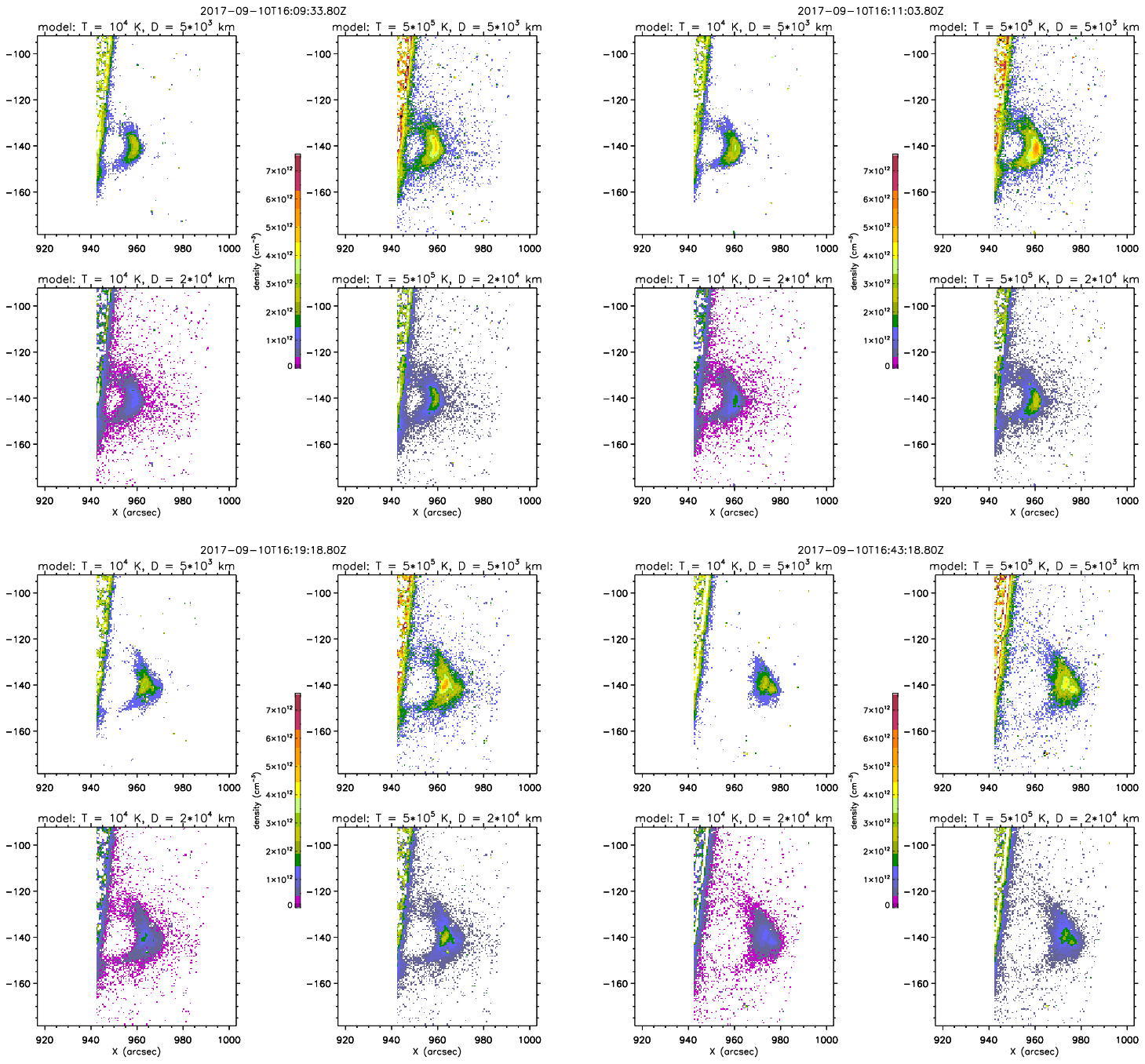}
\vspace{-8mm}
\caption{Temporal evolution of the electron density for four selected models computed from Equation~(\ref{e-all}) for four selected time steps. Units in $x$ and $y$ directions are in arcsec. It is visible that the maximum density is reached in the middle of the loop top.
A movie corresponding to this figure is available as online material of the journal.}
\label{f-density}
\end{figure*}

\subsection{Exploring the parameter space of electron densities}

From calibrated HMI WL intensity data we estimate the maximum radiation intensity in the flare loop system and the corresponding height above the solar surface 
(see Table~\ref{t-par}). Since we cannot obtain the loop temperature and effective thickness from SDO data, we computed the electron density by solving Equation~(\ref{e-all}) for a range of typical temperatures, taking into account cool as well as hot structures, for a grid of temperatures between 6000 and 10$^6$~K. For the effective thickness we take values between 200 and 20\,000~km. All together we have 208 models for a given time of the observations that allow us to explore the parameter space of the expected electron density. 

The electron density as a function of effective 
thickness and temperature is presented in the left panels of Figure~\ref{f-net} as contour plots for different times of the observations. Note that for the fourth selected time step we only focus on the left peak, which has a higher intensity (both peaks give rather similar values of electron density). The electron density is 
increasing with temperature and decreasing with effective 
thickness and it is more sensitive to effective thickness than temperature. For the maximal brightness at 16:11:03.8 UT and assuming effective thickness of 1000~km the electron density increases by a factor of two when comparing cool (6000~K) and hot loops (10$^6$~K). 
Arcades of cool loops would have 
$n_{\rm e}$ $\sim$ 7.3 $\times$ 10$^{12}$~cm$^{-3}$ while arcades of hot loops would have $n_{\rm e}$ $\sim$ 1.5 $\times$ 10$^{13}$~cm$^{-3}$. 
This behavior follows directly from Figure~\ref{f-ica}.
Normally the system of flare loops is a mixture of hot and cool loops  along the line of sight, thus the electron density is expected to lie between these two extreme values. We thus obtain a relatively high electron density of the order of 10$^{13}$~cm$^{-3}$ at $D_{\rm eff}$ = 1000~km. 

At higher effective thicknesses and for the same brightness, the obtained electron density is decreased. For example at $D_{\rm eff}$ = 10\,000~km, the electron density would be roughly between 
2.2 $\times$ 10$^{12}$~cm$^{-3}$ and 
4.5 $\times$ 10$^{12}$~cm$^{-3}$ for two extreme temperatures (by neglecting the Thomson scattering, the WL intensity would be proportional to emission measure n$_{\rm e}^2$ D$_{\rm eff}$). 

The right panels of Figure~\ref{f-net} show histograms of electron density at different times of the observations. The range of electron densities is higher at higher $I_{\rm WL}$
and is between 7.9 $\times$ 10$^{11}$ and 3.4 $\times$ 10$^{13}$~cm$^{-3}$ for our assumed parameter space and the maximum of the cut through the loop at our selected pixel. The mean weighted electron density is 3.8 $\times$ 10$^{12}$, 7.6 $\times$ 10$^{12}$, 7.0 $\times$ 10$^{12}$, 
and 4.1 $\times$ 10$^{12}$~cm$^{-3}$ for the four histograms.
To check the quality of the inversion, we can compute the WL radiation intensity from Equation~(\ref{e-all}), where the input parameters are temperature and electron density shown in Figure~\ref{f-net}. The difference between computed and observed $I_{\rm WL}$ is below 0.6~\%. 
For a comparison we also computed the ratio of the Thomson contribution to the total WL radiation intensity. The results are shown as histograms for all four 
different times of the observations in Figure~\ref{f-his}. The histograms clearly show that the Thomson contribution is relatively small, up to 15\% for low
WL radiation intensities and up to 5 \% for higher ones.

\subsection{Converting observed intensity maps into density maps}

Figure~\ref{f-density} shows the temporal evolution of the electron density for the two representative temperatures 
10$^4$ and 5 $\times$ 10$^5$~K and the two effective thicknesses 5000 and 20\,000~km by estimating the maximum radiation intensity at a height of 10\,000 km for four selected time steps. All other time steps can be found in the online movie. It is visible that the center of the loop top has the highest density, while it decreases towards the limb.

\section{Discussion and conclusions}
	\label{s-con}

In this paper we reported on SDO/HMI off-limb observations of a large X8.2 class flare, which was one of the strongest flares detected during the solar cycle 24. Well visible flare loops were seen
in the HMI pseudo-continuum channel during the gradual phase. Although similar WL loops have already been
analyzed for weaker flares \citep{oliv14,shil14}, this event is quite interesting due to its extraordinary brightness.

It is the first time that the HMI WL loop brightness is analyzed using  quantitative modeling, which includes all relevant emission processes. We demonstrate that for this strong flare the HMI intensities are dominated by the hydrogen recombination continuum, i.e. the Paschen continuum at the HMI wavelength 6173 \AA, with a small contribution
due to the tail of the Brackett continuum if assuming temperatures around 10$^4$ K. However, because we clearly see the multi-thermal character of the whole loop arcade from SDO/AIA imaging, we also consider the free-free emission, which plays a significant role at higher temperatures. Both the hydrogen recombination and the free-free emission are proportional to the loop emission measure, i.e. to the square of the mean electron
density. On the other hand, Thomson scattering of the photospheric light on the loop electrons is linearly proportional
to electron density and cannot explain the observed brightness. We show that the contribution of the Thomson scattering
to total continuum intensity is only a few percent at most in these very bright loops, but nevertheless we take it into
account when solving the quadratic Equation~(\ref{e-all}) for the electron density. 

The densities we obtain are unusually high for
flare loops in the gradual phase, ranging between 10$^{12}$ and 10$^{13}$ cm$^{-3}$ and mainly depending on the estimate of the
line of sight extension of the loop arcade. 

As shown by \citet{shil14} in case of their weaker flare, the Thomson-scattered radiation is partially linearly polarized and this was detected by HMI during their analyzed flare. In case of strong flares, the ratio of linear polarization $Q/I$ will be small, because $Q$ increases only linearly with the electron density while $I$, dominated by thermal processes (recombination and free-free emission) scales
quadratically with $n_{\rm e}$. For our observations, one would therefore not expect significant linear polarization, at least not co-spatial to the loop top where the density is high.
As suggested in \citet{shil14}, the particular distinction between Thomson scattering and processes proportional to the emission measure can be used to an advantage for an efficient disentangling between $n_{\rm e}$ and $D_{\rm eff}$. An analogous analysis method was developed for solar prominences \citep{jej09}, which also represent cool off-limb structures (note that flare loops have been previously classified as 'loop prominences', but now they are often called 'coronal rain',see e.g. \citeauthor{scu16} \citeyear{scu16}), however, the electron densities of prominences are low and thus the Thomson scattering completely dominates their WL emission, which can be detected
only during solar eclipses. In Figure~\ref{f-ica} $n_{\rm e}$= 10$^{11}$~cm$^{-3}$ refers to an upper limit of electron density usually met in quiescent prominences and the
continuum intensity is thus a few orders of magnitude lower than that detected in our studied flare loops. This also clearly explains why typical solar prominences have never been seen by HMI - their predicted intensity is apparently well below the detection limit.

In a next study we plan to analyze other HMI off-limb observations and derive the flare-loop electron densities 
which may help our understanding of the significance of WL loops on other flaring stars
and namely on those producing superflares as suggested by \citet{hei18}. The range of electron densities for a particular loop system can also be constrained by a detailed analysis of the linear-polarization signal from HMI. Moreover, for this X8.2 flare
spectral line data exists from various ground-based observations and these, together with complex non-LTE modeling
of flare loops, can also provide independent density diagnostics needed for a better understanding of evaporative processes and subsequent flare-loop cooling.

\begin{acknowledgements}
SJ acknowledges the financial support from the Slovenian Research Agency No. P1-0188. SJ, and PH acknowledge the support from the Czech Funding Agency through the grant No. 16-18495S and PH the partial
support from the grant No. 16-16861S. The funding from RVO-67985815 is also acknowledged. 
\end{acknowledgements}

\bibliographystyle{aasjournal}
\bibliography{biblio}


\end{document}